\documentclass[final,5p,twocolumn,times]{elsarticle}

\usepackage{ulem}
\usepackage{CJKutf8}
\usepackage{graphicx}
\usepackage{multirow}
\usepackage{subfigure}
\usepackage{verbatim}
\usepackage{multirow}
\usepackage{booktabs}
\usepackage{siunitx}

\newcommand{\ra}[1]{\renewcommand{\arraystretch}{#1}}

\begin{document}

\begin{frontmatter}
\title{Symmetry energy investigation with pion production from Sn+Sn systems}

\author[nscl]{G.~Jhang\corref{cor}} \ead{changj@nscl.msu.edu}
\author[nscl,msu]{J.~Estee}
\address[nscl]{National Superconducting Cyclotron Laboratory, Michigan State University, East Lansing, Michigan 48824, USA}
\address[msu]{Department of Physics, Michigan State University, East Lansing, Michigan 48824, USA}
\author[nscl,msu]{J.~Barney}
\author[nscl]{G.~Cerizza}
\author[riken,kyoto]{M.~Kaneko}
\address[riken]{RIKEN Nishina Center, Hirosawa 2-1, Wako, Saitama 351-0198, Japan}
\address[kyoto]{Department of Physics, Kyoto University, Kita-shirakawa, Kyoto 606-8502, Japan}
\author[korea]{J.~W.~Lee}
\address[korea]{Department of Physics, Korea University, Seoul 02841, Republic of Korea}
\author[nscl,msu]{W.~G.~Lynch\corref{cor}} \ead{lynch@nscl.msu.edu}
\author[riken]{T.~Isobe\corref{cor}} \ead{isobe@riken.jp}
\author[riken]{M.~Kurata-Nishimura}
\author[kyoto]{T.~Murakami\corref{cor}} \ead{murakami.tetsuya.3e@kyoto-u.ac.jp}
\author[nscl,msu]{C.~Y~.Tsang}
\author[nscl,msu]{M.~B.~Tsang\corref{cor}} \ead{tsang@nscl.msu.edu}
\author[nscl]{R.~Wang}
\author[riken]{D.~S.~Ahn}
\author[itud,gsi]{L.~Atar}
\address[itud]{Institut f\"ur Kernphysik, Technische Universit\"at Darmstadt, D-64289 Darmstadt, Germany}
\address[gsi]{GSI Helmholtzzentrum f\"ur Schwerionenforschung, Planckstrasse 1, 64291 Darmstadt, Germany}
\author[itud,gsi]{T.~Aumann}
\author[riken]{H.~Baba}
\author[gsi]{K.~Boretzky}
\author[juk]{J.~Brzychczyk}
\address[juk]{Faculty of Physics, Astronomy and Applied Computer Science, Jagiellonian University, Krak\'ow, Poland}
\author[riken]{N.~Chiga}
\author[riken]{N.~Fukuda}
\author[rbi,riken]{I.~Gasparic}
\address[rbi]{Division of Experimental Physics, Rudjer Boskovic Institute, Zagreb, Croatia}
\author[korea]{B.~Hong}
\author[itud,gsi]{A.~Horvat}
\author[rikkyo]{K.~Ieki}
\address[rikkyo]{Department of Physics, Rikkyo University, Nishi-Ikebukuro 3-34-1, Tokyo 171-8501, Japan}
\author[riken]{N.~Inabe}
\author[ibs]{Y.~J.~Kim}
\address[ibs]{Rare Isotope Science Project, Institute for Basic Science, Daejeon 34047, Republic of Korea}
\author[tohoku]{T.~Kobayashi}
\address[tohoku]{Department of Physics, Tohoku University, Sendai 980-8578, Japan}
\author[titech]{Y.~Kondo}
\address[titech]{Department of Physics, Tokyo Institute of Technology, Tokyo 152-8551, Japan}
\author[inp]{P.~Lasko}
\address[inp]{Institute of Nuclear Physics PAN, ul. Radzikowskiego 152, 31-342 Krak\'ow, Poland}
\author[ibs]{H.~S.~Lee}
\author[gsi]{Y.~Leifels}
\author[inp]{J.~Łukasik}
\author[nscl,msu]{J.~Manfredi}
\author[tamu]{A.~B.~McIntosh}
\address[tamu]{Cyclotron Institute, Texas A\&M University, College Station, Texas 77843, USA}
\author[nscl]{P.~Morfouace}
\author[titech]{T.~Nakamura}
\author[riken,kyoto]{N.~Nakatsuka}
\author[riken]{S.~Nishimura}
\author[tamu]{R.~Olsen}
\author[riken]{H.~Otsu}
\author[inp]{P.~Paw\l{}owski}
\author[juk]{K.~Pelczar}
\author[itud]{D.~Rossi}
\author[riken]{H.~Sakurai}
\author[nscl]{C.~Santamaria}
\author[riken]{H.~Sato}
\author[itud]{H.~Scheit}
\author[nscl]{R.~Shane}
\author[riken]{Y.~Shimizu}
\author[gsi]{H.~Simon}
\author[nnisp]{A.~Snoch}
\address[nnisp]{Nikhef National Institute for Subatomic Physics, Amsterdam, Netherlands}
\author[juk]{A.~Sochocka}
\author[juk]{Z.~Sosin\fnref{dec}}
\fntext[dec]{Deceased}
\author[riken]{T.~Sumikama}
\author[riken]{H.~Suzuki}
\author[riken]{D.~Suzuki}
\author[riken]{H.~Takeda}
\author[nscl]{S.~Tangwancharoen}
\author[itud,gsi]{H.~Toernqvist}
\author[rikkyo]{Y.~Togano}
\author[tu]{Z.~G.~Xiao}
\address[tu]{Department of Physics, Tsinghua University, Beijing 100084, PR China}
\author[tamu,dctamu]{S.~J.~Yennello}
\address[dctamu]{Department of Chemistry, Texas A\&M University, College Station, Texas 77843, USA}
\author[nscl]{J.~Yurkon}
\author[tu]{Y.~Zhang}
\author{\\the S$\pi$RIT Collaboration}

\author[mc]{\\Maria~Colonna}
\address[mc]{INFN-LNS, Laboratori Nazionali del Sud, 95123 Catania, Italy}
\author[dc]{Dan~Cozma}
\address[dc]{IFIN-HH, Reactorului 30, 077125 M\v{a}gurele-Bucharest, Romania}
\author[nscl,msu]{Pawe\l{}~Danielewicz}
\author[gsi,itp,fias]{Hannah~Elfner}
\address[itp]{Institute for Theoretical Physics, Goethe University, 60438 Frankfurt am Main, Germany}
\address[fias]{Frankfurt Institute for Advanced Studies, Johann Wolfgang Goethe University, 60438 Frankfurt am Main, Germany}
\author[tottori]{Natsumi~Ikeno}
\address[tottori]{Department of Life and Environmental Agricultural Sciences, Tottori University, Tottori 680-8551, Japan}
\author[tamu,dpatamu]{Che~Ming~Ko}
\address[dpatamu]{Department of Physics and Astronomy, Texas A\&M University, College Station, Texas 77843, USA}
\author[gsi,itp,fias]{Justin~Mohs}
\author[llbl]{Dmytro~Oliinychenko}
\address[llbl]{Lawrence Berkeley Laboratory, 1 Cyclotron Road, Berkeley, California 94720, USA }
\author[tohoku]{Akira~Ono}
\address[tohoku]{Department of Physics, Tohoku University, Sendai 980-8578, Japan}
\author[sfinet]{Jun~Su}
\author[yjw]{Yong~Jia~Wang}
\address[yjw]{School of Science, Huzhou University, Huzhou 313000, China}
\author[hw]{Hermann~Wolter}
\address[hw]{Faculty of Physics, Ludwig-Maximilians-University of Munich, D-85748 Garching, Germany}
\author[jx1,jx2]{Jun~Xu}
\address[jx1]{Shanghai Advanced Research Institute, Chinese Academy of Sciences, Shanghai 201210, China}
\address[jx2]{Shanghai Institute of Applied Physics, Chinese Academy of Sciences, Shanghai 201800}
\address[ciae]{China Institute of Atomic Energy, Beijing 102413, China}
\author[ciae]{Ying-Xun~Zhang}
\author[sfinet]{Zhen~Zhang}
\address[sfinet]{Sino-French Institute of Nuclear Engineering and Technology, Sun Yat-sen University, Zhuhai 519082, China}
\author{\\the TMEP collaboration}

\cortext[cor]{Corresponding authors}

\begin{abstract}
In the past two decades, pions created in the high density regions of heavy ion collisions have been predicted to be sensitive at high densities to the symmetry energy term in the nuclear equation of state, a property that is key to our understanding of neutron stars. In a new experiment designed to study the symmetry energy, the multiplicities of negatively and positively charged pions have been measured with high accuracy for central $^{132}$Sn+$^{124}$Sn, $^{112}$Sn+$^{124}$Sn, and $^{108}$Sn+$^{112}$Sn collisions at $E/A=\SI{270}{MeV}$ with the S$\pi$RIT Time Projection Chamber. While the uncertainties of individual pion multiplicities are measured to 4\%, those of the charged pion multiplicity ratios are measured to 2\%. We compare these data to predictions from seven major transport models.  The calculations reproduce qualitatively the dependence of the multiplicities and their ratios on the total neutron to proton number in the colliding systems. However, the predictions of the transport models from different codes differ too much to allow extraction of reliable constraints on the symmetry energy from the data. This finding may explain previous contradictory conclusions on symmetry energy constraints obtained from pion data in Au+Au system. These new results call for better understanding of the differences among transport codes, and new observables that are more sensitive to the density dependence of the symmetry energy.   
\end{abstract}
\end{frontmatter}

\section{Introduction}

Gravitational waves (GW) from the first binary neutron star merger event GW170817 observed by the LIGO-VIRGO collaboration have provided a glimpse into the properties of asymmetric compact nuclear objects with large imbalance of protons and neutrons under extreme conditions \cite{Abb17, Abb18}. The eventual fate of such merged objects as giant neutron stars or as transient neutron stars that later collapse into black holes are currently not known \cite{Abb19}. It depends on the equation of state (EoS) of very neutron-rich nuclear matter that is of great interest to astronomy and astrophysics \cite{Pir19, Rad19}. In  nuclear physics \cite{Cai15, Hor14}, understanding the nuclear EoS has motivated research in dense matter and the development of new powerful rare isotope accelerator facilities worldwide.

Currently, astrophysical observations and nuclear physics experiments provide complementary information about the nuclear EoS \cite{Abb18, Tsa19b, Ste13, Dan02}.  Measurements of nucleus-nucleus collisions and their interpretations via transport models have provided independent and consistent constraints on the EoS of symmetric matter \cite{Dan02,Lyn09,LeF16} which has equal numbers of neutrons and protons.  By combining such laboratory constraints with the GW results, the density dependence of the symmetry pressure, which contributes when neutron and proton densities differ, has been obtained with large uncertainties for densities above the saturation density, i.e. $\rho_0=\SI{1.74e14}{g/cm^3}$ \cite{Tsa19b}.

   Neutron star calculations have shown the deformability of 1.4 solar mass neutron stars \cite{Tsa19, Lim18, Tews18} to be strongly correlated with the symmetry pressure contributions to the nuclear matter EoS at twice saturation density in the outer core of the neutron star. The symmetry pressure helps support the star against the gravitation force while the symmetry energy plays a dominant role in the internal structure of neutron stars \cite{Cai15, Lat01, Dra14, Duc11, Lat16}. In the past two decades, neutron star calculations have predicted that the presence of pions and $\Delta$  isobars \cite{Dra14,Jli19} may strongly influence the symmetry pressure at such densities and the thermal properties and post merger dynamics of neutron stars as well \cite{For20}. To address questions concerning the microscopic foundation of the EoS in neutron stars, laboratory constraints on the symmetry energy or pressure and the role of pions at supra-saturation densities are needed.

Since the symmetry energy governs the dependence of the EoS on neutron excess, one can study nuclear collisions in laboratory using beams with extreme neutron richness impinging on targets composed of neutron-rich or neutron-deficient nuclei.  Compared to neutron stars, the symmetry energy plays a much smaller role in nuclear collisions, reflecting the smaller asymmetries of atomic nuclei. To maximize the sensitivity to the symmetry energy and minimize sensitivities to other quantities, one can employ isovector observables such as the relative emission of isospin multiplets, e.g. $\pi^-$ vs. $\pi^+$, $n$ vs. $p$, $t$ vs. $^3$He, etc., that are subject to symmetry forces of opposite sign \cite{Che05, Cou16, Li08, Tsa04, Tsa09, Mor19}. 

Here, we report the results from the first measurements dedicated to probe the symmetry energy via pion production in Sn+Sn collisions. 
We designed Our experiment is designed to measure pion measure very low momentum pions, to ensure accurate measurements of pion energy spectra and pion multiplicities. Because negatively and positively charged pions ($\pi^-$ and $\pi^+$) are primarily produced in n-n collisions and p-p collisions, respectively, pion yield ratios $Y(\pi^-)/Y(\pi^+)$ from central collisions are predicted to be sensitive to the isovector mean field potentials that contribute to the symmetry energy at high densities \cite{Hon14, Ike16, Ike20, Li02}. To retain sensitivity to the symmetry energy while suppressing the influence of the Coulomb interaction, we construct double ratios, i.e., ratios of pion yield ratios for two different reactions with the same total charge but very different isospin asymmetries. 

Transport models are required to extract constraints on the EoS from heavy-ion collisions and have provided constraints on the symmetric matter EoS and its associated isoscalar mean field potentials \cite{Dan02,LeF16}. In transport models, pions are produced by excitation and decay of the $\Delta$ resonance in the compressed high density region formed during the early stages of a nucleus-nucleus collision \cite{Hon14, Ike16, Ike20, Li02}. Recent efforts to constrain the isovector mean fields and the associated symmetry energy, based on pion multiplicities from the Au+Au data have led to inconsistent conclusions \cite{Fen10, Xia09, Xie13}. To understand the discrepencies among codes, the Transport Model Evaluation Project (TMEP), has formulated benchmark calculations to evaluate methods used to solve transport equations and predict experimental observables \cite{Ono19, Xu16, Zha18}. The latest publication \cite{Ono19} discusses transport calculations of pion production in a box with periodic boundary conditions where analytic solutions are known. When identical input parameters are used, good agreement, especially of the pion yield ratios, was obtained for most codes \cite{Ono19}.  

\begin{figure}[ht]
    \centering
    \includegraphics[width=\linewidth]{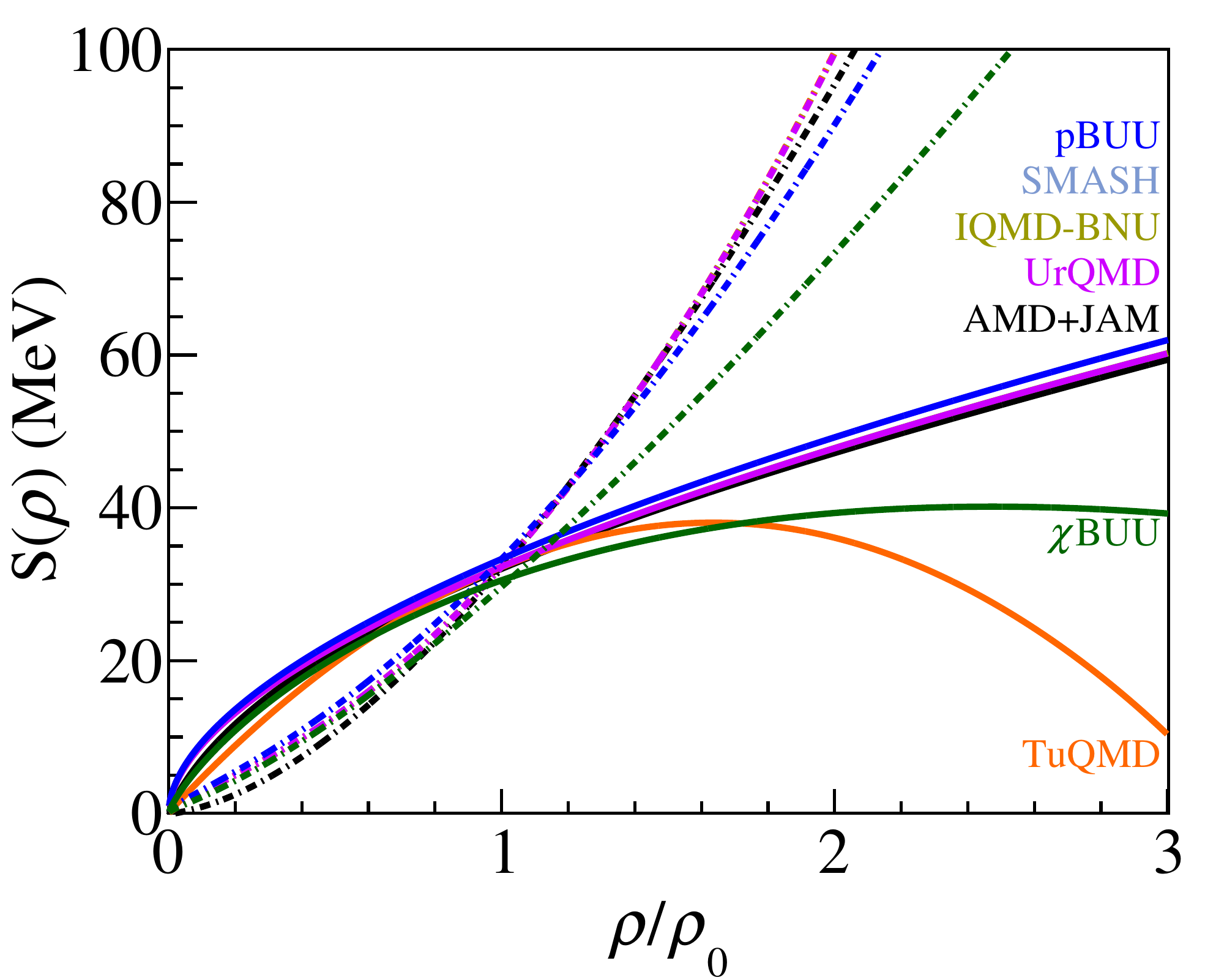}
    \caption{Nuclear symmetry energies used in the seven transport models to study pion production in Sn+Sn reactions. The solid and dash-dotted lines of the same color correspond to the soft and stiff symmetry energies, respectively, for each code. Some codes use the same EoS and cannot be seen: the soft density dependence of the symmetry energy for SMASH, IQMD-BNU are the same as those used in AMD+JAM;  Similarly, TuQMD employs the same stiff density dependent symmetry energy as UrQMD and SMASH. These three curves (not shown) are nearly the same as the one used by AMD+JAM.}
    \label{eos}
\end{figure}

In the following we show seven transport model predictions made without knowledge of the present experimental data, These seven widely used transport codes are: (i) AMD+JAM \cite{Ike16, Ike20, Nar99}, (ii) IQMD-BNU \cite{Su11, Su13, Su14}, (iii) pBUU \cite{Hon14, Dan00}, (iv) SMASH \cite{Wei16}, (v) TuQMD \cite{Coz16,Coz17},  (vi) UrQMD \cite{Ba98, Li11} and (vii) $\chi$BUU \cite{Zha18k} which is a variant of RVUU \cite{Ko87, Ko88} using the Sk$\chi$m$^{*}$ energy functional \cite{Lim17}. The above codes fall into two categories with the $\chi$BUU, SMASH and pBUU based on the Boltzmann-Uehling-Ulhenbeck equation, and those based on the quantum molecular dynamics, includes the TuQMD, AMD+JAM, IQMD-BNU, and UrQMD. The differences between the BUU and QMD code families, as well as details of the codes have been described in the published code evaluation studies~\cite{Ono19, Xu16, Zha18}, especially in Ref.~\cite{Zha18}. These codes are listed in Table~\ref{tab:pionYields} together with their results for the pion multiplicities and ratios, which will be discussed below. Each code performed calculations using two extreme choices of the isovector nucleon mean field potentials characterized in the second column of Table~\ref{tab:pionYields} by $L$, the slope of the symmetry energy at saturation. The  density dependence of the symmetry energy used by the codes (identified by color and labels) is displayed in Fig.~\ref{eos}. For clarity, we loosely refer to the symmetry energy dependence as stiff (dot dashed curves in the top two groups) and soft (solid curves in the bottom two groups of curves). Due to overlapping $L$ values, some lines cannot be displayed separately as described in the caption. 
In this paper we will demonstrate the present uncertainties or spread in the predictions and the sensitivity of each code to the symmetry energy. 
 
\section{Experimental Setup}

Four reactions were measured covering a wide range of asymmetries characterized by the neutron to proton ratios, $N/Z$: (a) $^{132}$Sn$+^{124}$Sn ($N/Z=1.56$), (b) $^{108}$Sn$+^{112}$Sn ($N/Z=1.2$), (c) $^{112}$Sn$+^{124}$Sn ($N/Z=1.36$), and (d) $^{124}$Sn$+^{112}$Sn ($N/Z=1.36$), at the Radioactive Isotope Beam Factory operated by the RIKEN Nishina center and CNS, University of Tokyo. We collided secondary beams of $^{132}$Sn, $^{124}$Sn, $^{112}$Sn, and $^{108}$Sn at $E/A=270$~MeV onto isotopically enriched ($>~95\%$) $^{112}$Sn and $^{124}$Sn targets of 561 and 608~mg/cm$^2$ areal density, respectively. The low purity ($\sim10\%$) of the $^{124}$Sn beam relative to the much higher purity $>50\%$ of $^{108,112,132}$Sn beams complicates the inclusion of $^{124}$Sn-induced reactions in the current analyses. 

To measure charged pions and light isotopes with Z$\le$3 over the required experimental acceptance, we built the SAMURAI pion Reconstruction Ion Tracker (S$\pi$RIT) Time Projection Chamber (TPC) and the associated trigger arrays \cite{Bar20, Las17, Ots16, Sha15}. Detailed description of experimental setup and performance of the S$\pi$RIT TPC can be found in Refs.~\cite{Bar20, Bar19}. Technical challenges of various aspects of the experiments including software techniques used in data analysis have been documented in at least 9 publications \cite{Bar20, Las17, Sha15, Est19, Iso18, Jha16, Lee20, Tan17, Tsa20}.   In this letter, only relevant information about the experimental setup and analysis regarding the pion measurements are briefly summarized.  This letter is the first of a series of physics results obtained from the S$\pi$RIT experiments. Additional results will be published as the analysis is finalized.

\begin{figure}[ht]
    \centering
    \includegraphics[width=\linewidth]{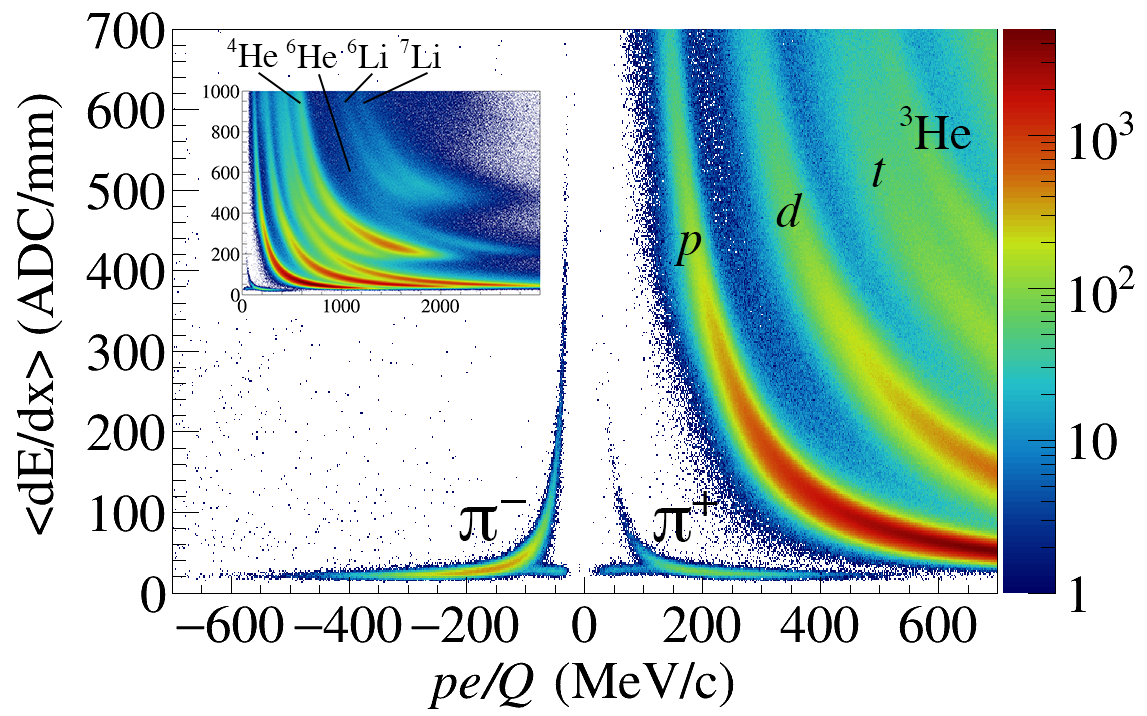}
    \caption{Particle identification plot from the $^{132}$Sn +$^{124}$Sn reaction, measured with the S$\pi$RIT TPC. The main plot focuses on the resolution of the pion; the broader spectrum of positively charged particles is shown in the inset.}
    \label{fig2}
\end{figure}

This paper mainly focuses on the measurement of charged pions. Charged particles detected in each event is used to infer the impact parameters. Even though a new electronic system, the Generic Electronics for TPCs (GET), was employed to attain a large dynamic range when measuring the energy loss signals in the TPC \cite{Iso18}, novel software techniques were needed to achieve an effective signal to noise ratio  of 4000 to 1 \cite{Est19}. This allowed good isotope separation of pions and light charged nuclei up to Li as shown in  Fig.~\ref{fig2}. A tracking analysis framework called S$\pi$RITROOT is created specifically to reconstruct the momentum and energy losses of each particle track~\cite{Lee20, Github}. It can also interface with Geant4 toolkits to simulate the TPC response for the Monte Carlo (MC) tracks. In this way, the same analysis software evaluates both the detected events and MC tracks. The efficiency of the TPC was estimated by embedding MC tracks into real experimental data \cite{Abe09}. 

\section{Results}

Assuming that the measured charged particle multiplicity decreases monotonically with increasing impact parameter, we select events with the highest multiplicities while retaining good statistical accuracy. The selected events have a cross section of about 0.3~barns, corresponding to a mean impact parameter of $\approx3$~fm \cite{Bar20}. Fig.~\ref{fig2} shows the charged particle identification achieved for the $^{132}$Sn +$^{124}$Sn reaction by combining the measured magnetic rigidity ($pe/Q$) and the energy loss ($dE/dx$) in the counter gas of the S$\pi$RIT TPC. The pions are located at very low $dE/dx$ regions. Electrons and positrons from the Dalitz decay of $\pi^0$ are the largest contributions in the pion background. They appear as horizontal appendages to the pion PID at low rigidity. These background contributions are insignificant. Nonetheless, they are subtracted via methods detailed in Ref.~\cite{Bar19}.
        
The design of the S$\pi$IRIT experiment allows transverse momentum of pions to be measured down to 0 MeV/c in the center-of-mass (CM) system, the key to obtain accurate pion multiplicities and their ratios. In contrast, previous measurements of Au+Au system had thresholds of $\approx\SI{100}{MeV/c}$ \cite{Rei07}. We define the angles in a coordinate system, where the z-axis lies along the beam line, y-axis is vertical pointing upwards and x-axis lies in the horizontal plane conforming to the definition of a right-handed coordinate system. To ensure accurate pion momenta, we focus on pions measured to polar angles of $\theta_{CM}~<~\ang{90}$. In central collisions, pion emission is azimuthally symmetric; thus our pion angular coverage at rapidities $y>y_{CM}$ is complete. To obtain the total multiplicities, we further assume that the pion multiplicities at $y>y_{CM}$ and $y<y_{CM}$ are equal; an assumption supported by measuring the pion emission from the forward and reverse reactions of $^{124}$Sn +$^{112}$Sn and $^{112}$Sn +$^{124}$Sn systems \cite{Bar19}. Using pBUU, we estimate that the pion multiplicities from  $y>y_{CM}$ and $y<y_{CM}$, are the same to within 8\%. Given the uncertainties in code predictions described below, this small difference is not an issue of concern. Of course, exact comparisons to our data can always be made by comparing theoretical calculations at $y>y_{CM}$ . 
  
\begin{figure}
    \centering
    \includegraphics[width=0.8\linewidth]{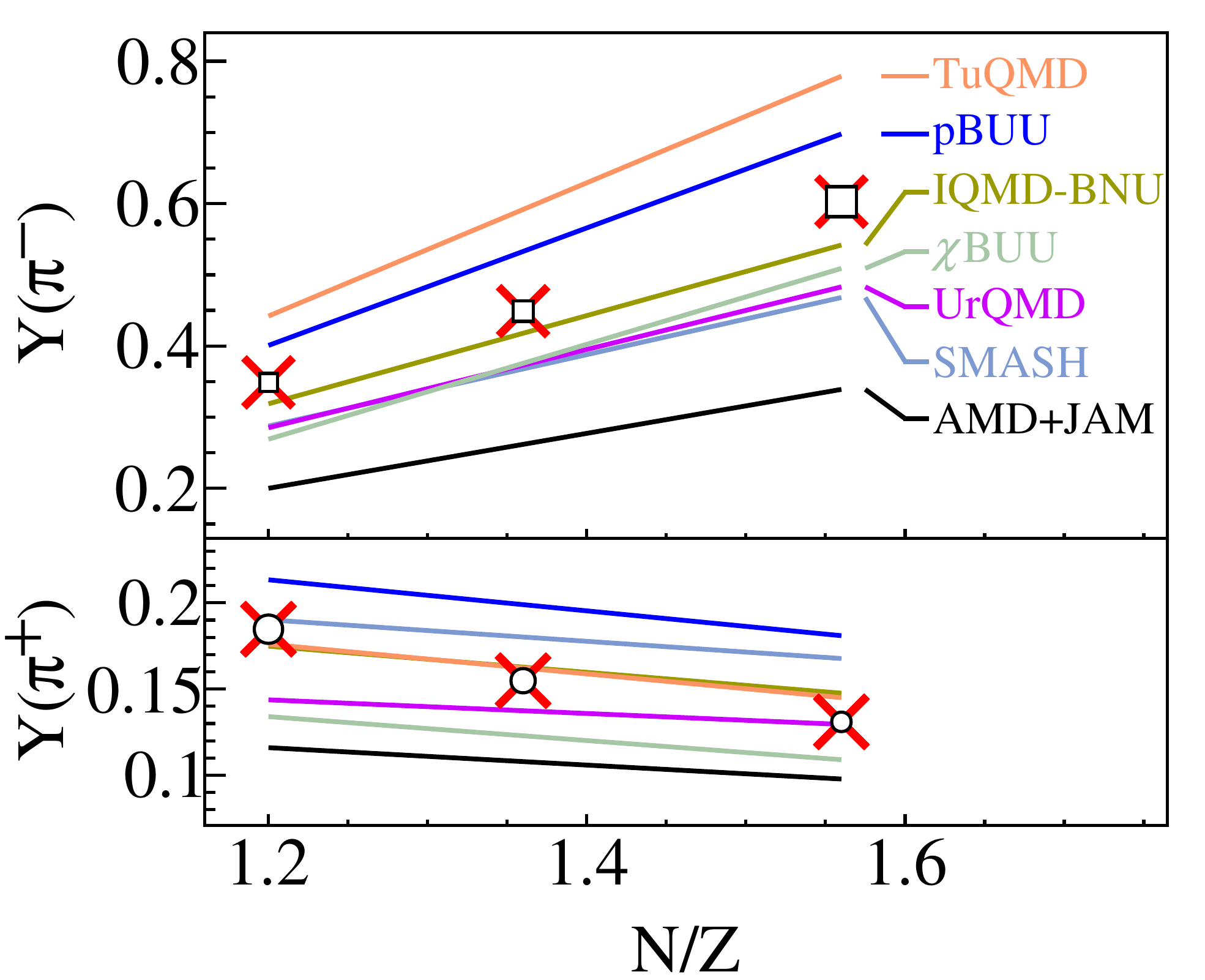}
    
    \caption{Charged pion multiplicities as a function of $N/Z$ for three reaction systems, $^{132}$Sn +$^{124}$Sn, $^{112}$Sn +$^{124}$Sn, and $^{108}$Sn +$^{112}$Sn, for $\pi^-$ and $\pi^+$ in the upper and lower panels, respectively. Crosses are the experimental data with uncertainties represented by the sizes of the open symbols. Lines  are the calculations of the codes identified by the legend and listed in Table~\ref{tab:pionYields} for the choice of the soft symmetry energy for the two systems with highest and lowest neutron content connected to guide the eye.}
    \label{fig3}
\end{figure}

The experimental results for the total pion multiplicities are shown in Fig.~\ref{fig3} together with the results of the calculations to be discussed below. The crosses overlaid with open symbols show the measured $\pi^-$ and  $\pi^+$ multiplicities as a function of $N/Z$, in the top and bottom panel, respectively, for the three systems $^{108}$Sn +$^{112}$Sn, $^{112}$Sn +$^{124}$Sn, and $^{132}$Sn +$^{124}$Sn.  The size of the open symbols at the center of the crosses corresponds to the combined experimental (systematic and statistical) uncertainties. The systematic errors associated mainly with experimental analysis \cite{Bar19} for the $\pi^-$ and $\pi^+$ multiplicities are 3\% and 4\%, respectively, with the statistical uncertainties less than 1.2\%. The $N/Z$ ratios of the initial system are chosen for the abscissa in Fig.~\ref{fig3}. A more interesting asymmetry variable might be the $N/Z$ ratios of the participant regions, which is model dependent.

\begin{figure*}
    \centering
    \includegraphics[width=0.7\linewidth]{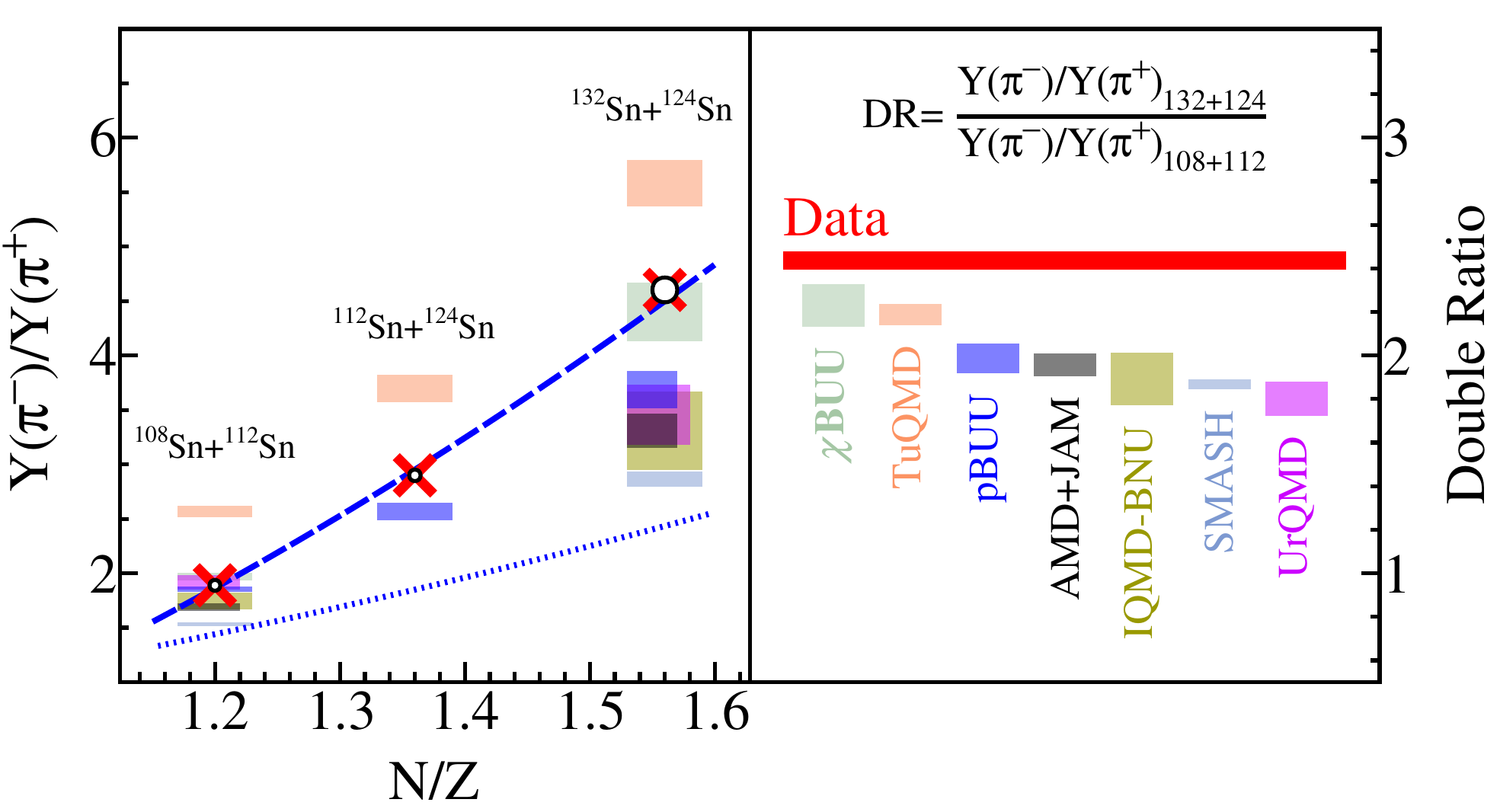}
    \caption{(Left panel) Charged pion yield ratios as a function of $N/Z$. The data are shown as crosses with the size of the open symbols representing the experimental errors. The results of the calculations are represented by colored boxes for the different codes identified by their color in the right panel. The height of the boxes is given by the difference of predictions for the soft and stiff symmetry energies. The dashed blue line is a power-law fit with the function $(N/Z)^{3.6}$, while the dotted blue line gives $(N/Z)^2$ of the system. (Right panel) Double pion yield ratios for $^{132}$Sn +$^{124}$Sn and $^{108}$Sn +$^{112}$Sn. The data and their uncertainty are given by the red horizontal bar and the results of the transport models are shown by the colored boxes, in a similar way as in the left panel. See text for details.}
    \label{ratios}
\end{figure*}

 The $\pi^-$ multiplicities increase while the $\pi^+$ multiplicities decrease with $N/Z$ (Fig.~\ref{fig3}). Consequently, the $\pi^-/\pi^+$ single ratios shown in Fig.~\ref{ratios} (left panel) increase  steeply with $N/Z$. As systematic errors common to both the $\pi^-$ and $\pi^+$ multiplicities cancel in the ratios; the overall errors, represented by the size of the open symbols are reduced to less than 2\%. The blue dashed curve that passes through the data is a power-law fit $(N/Z)^{3.6}$, while a $(N/Z)^2$ dependence (shown as a blue dotted line) would be expected from a $\Delta$ resonant model for pion production \cite{Rei07,Li02}. The measured $\pi^-/\pi^+$ ratios and many of the calculated ratios are considerably larger than the $(N/Z)^2$ of the system indicating that other dynamical factors beyond the simple $\Delta$ resonant model play a role here. 

The double pion yield ratios (DR) would show a more selective dependence on the symmetry energy \cite{Tsa17}. Only the largest double ratio value, $2.42 \pm 0.05$ from the two extreme reactions, $[Y(\pi^-)/Y(\pi^+)]_{132+124}/[Y(\pi^-)/Y(\pi^+)]_{108+112}$ is plotted as a horizontal bar in the right panel of Fig.~\ref{ratios}. Its uncertainty (2$\%$), represented by the vertical height of the bar, reflects a large cancellation of systematic errors. Achieving this experimental accuracy is important because most transport model calculations predict the impact of the symmetry energy on this observable to be less than 10\% \cite{Li08, Ike16, Ike20, Xia09}. 

\section{Transport Model Predictions}
\begin{table*}
\centering
\ra{1.1}
\begin{tabular}{@{}ccccccccccc@{}}
\toprule
        &  & \multicolumn{3}{c}{(a) $^{132}$Sn+$^{124}$Sn} &  & \multicolumn{3}{c}{(b) $^{108}$Sn+$^{112}$Sn} &  \\
        \cmidrule{3-5} \cmidrule{7-9} 
        Code name & $\mathrm{L~(MeV)}$ &  $\mathrm{Y}(\pi^-)$ & $\mathrm{Y}(\pi^+)$ & $\mathrm{SR}(\pi^-/\pi^+)$ &  & $\mathrm{Y}(\pi^-)$  & $\mathrm{Y}(\pi^+)$ & $\mathrm{SR}(\pi^-/\pi^+)$ &  $\mathrm{DR}(\pi^-/\pi^+)$ \\
    \midrule
        $\chi$BUU& 45.6 & 0.509 & 0.109 & 4.67 & & 0.269 & 0.134 & 2.01 & 2.33 \\
        & 120 & 0.483 & 0.117 & 4.13 &  & 0.271 & 0.140 & 1.94 & 2.13 \\
        \addlinespace[0.2cm]
        TuQMD & 54.6 & 0.779 & 0.145 & 5.37 &  & 0.442 & 0.176 & 2.51 & 2.14 \\
        & 145 & 0.839 & 0.145 & 5.79  &  & 0.474 & 0.181 & 2.62 & 2.21 \\
        \addlinespace[0.2cm]
        pBUU & 56.1 & 0.698 & 0.181 & 3.86 &  & 0.401 & 0.213 & 1.88 & 2.05 \\
        & 135 & 0.649 & 0.185 & 3.51  &   & 0.392 & 0.214 & 1.83 & 1.92 \\
        \addlinespace[0.2cm]
        AMD+JAM& 55 & 0.339 & 0.0978 & 3.47 &  & 0.200 & 0.116 & 1.72 & 2.02 \\
        & 152 & 0.311 & 0.0986 & 3.15 &  & 0.192 & 0.116 & 1.66 & 1.90 \\
        \addlinespace[0.2cm]
        IQMD-BNU & 54.6 & 0.542 & 0.148 & 3.67 &  & 0.319 & 0.175 & 1.82 & 2.01 \\
        & 145 & 0.452 & 0.153 & 2.95 &  & 0.278 & 0.167 & 1.67 & 1.77 \\
        \addlinespace[0.2cm]
        SMASH  & 55 & 0.468 & 0.168 & 2.79  &  & 0.287 & 0.190 & 1.51 & 1.85 \\
        & 152 & 0.479 & 0.163 & 2.93 &  & 0.292 & 0.188 & 1.55 & 1.89 \\
        \addlinespace[0.2cm]
        UrQMD & 54 & 0.483 & 0.129 & 3.73 &  & 0.285 & 0.144 & 1.98 & 1.88 \\
        & 144 & 0.447 & 0.141 & 3.18 &  & 0.275 & 0.149 & 1.85 & 1.72 \\
        \addlinespace[0.2cm]
        Data & -- & 0.603(20) & 0.131(5) & 4.60(11) &  & 0.349(12) & 0.185(8) & 1.89(4) & 2.44(4) \\
        \bottomrule
    \end{tabular}
    \caption{Pion multiplicities $\mathrm{Y}(\pi^{\pm})$, single multiplicity ratios $\mathrm{SR}(\pi^-/\pi^+)$ and double multiplicity ratios $\mathrm{DR}(\pi^-/\pi^+)$ from seven transport codes and compared to the data (last row). Each code uses two different symmetry energy functionals, characterized in the second column by the slope $L$ of the symmetry energy at saturation, and shown in their full density dependence in Fig.~\ref{eos}. Experimental errors include both systematic and statistical uncertainties. For most calculations, the statistical errors are negligibly small ($<$0.5\%) as they are obtained with a sufficient number of events in the simulations. The present study thus indicates that the theoretical uncertainties due to disagreement of codes are very large. }
    \label{tab:pionYields}
\end{table*}

Next we confront these experimental data with predictions on the pion production by seven widely used transport codes listed in Table~\ref{tab:pionYields}, using two choices for the stiffness of the symmetry energy, shown in Fig.~\ref{eos}. All other physical input to the codes, like the isoscalar mean fields and the elastic and inelastic
cross sections, are chosen by the code authors according to their present best modeling of heavy-ion collisions, and are not common to all codes. Unlike earlier theoretical analyses of the Au+Au data, the calculations shown here are actual predictions for this experiment based on the best effort of the respective codes. No model parameters have been adjusted.
The calculations were requested for the two reactions of extreme isospin content, $^{132}$Sn +$^{124}$Sn and $^{108}$Sn +$^{112}$Sn at $\langle b\rangle= \SI{3}{fm}$. 

The results of the codes, listed in Table~\ref{tab:pionYields}, are shown graphically in  Fig.~\ref{fig3} for the $\pi^-$ and $\pi^+$ yields in the upper and lower panels respectively, and in Fig.~\ref{ratios} for the single and double ratios. The lines in  Fig.~\ref{fig3}  represent the yields of $\pi^-$ and $\pi^+$ using the soft density dependent symmetry energy for each code. The results using the stiff symmetry energy are not plotted in Fig.~\ref{fig3} for clarity. The variation of the predicted $\pi^+$ multiplicities in the lower panel is smaller than that for the $\pi^-$ multiplicities in the upper panel. All the codes (except for SMASH) reproduce the general trends of the $N/Z$ dependence of the data i.e. larger than the expected $(N/Z)^2$ dependence. However, there are large differences among code predictions, often larger than their deviation from the data.

The left panel of Fig.~\ref{ratios} shows the measured and predicted $\mathrm{Y}(\pi^-)/\mathrm{Y}(\pi^+)$ yield ratios. Predictions for the $^{112}$Sn +$^{124}$Sn reaction are shown only for TuQMD and pBUU codes, as the original assignment for model predictions did not include this system. The results for both choices of the symmetry energy are shown by colored boxes, where the upper and lower borders of each box correspond to the results from the stiff or the soft symmetry energies. All codes predict sensitivity to the symmetry energy even though the degrees of sensitivity (height of the boxes) is generally small in relation to the difference to the data. In most cases the soft symmetry energy results in the higher ratio, except for the codes TuQMD and SMASH, for which the opposite holds true (see Table~\ref{tab:pionYields}). The differences among predictions for the single ratios are smaller than for the individual multiplicities. However, they are still larger than the overall differences between the calculations and the experimental data. 

The $N/Z$ dependence of these predictions agree qualitatively, but not quantitatively with the data. This suggests that other factors such as specific details in the $\Delta$ production cross-sections, isoscalar mean-field potential, the accuracy in the treatment of the Coulomb interaction or differences in the magnitude of neutron skins of the initial state nuclei could indirectly influence the single ratios. To reduce the sensitivity to such effects, we construct the double ratio, $[Y(\pi^-)/Y(\pi^+)]_{132+124}/[Y(\pi^-)/Y(\pi^+)]_{108+112}$. Since all reactions have the same total charge, the double ratio largely removes the Coulomb effects and any multiplicative normalization problem with the $\Delta$ and pion ($\pi^+$ or $\pi^-$) sub-threshold production cross sections. The theoretical predictions for the double ratio are plotted as rectangular boxes in the right panel of Fig.~\ref{ratios} with borders corresponding to the soft and the stiff symmetry energies. The red horizontal bar represents the experimental value.  
The best agreement with the double ratio data is provided by the $\chi$BUU \cite{Zha18k} and TuQMD. Although both models include threshold effects on $\Delta$ resonance production, they predict opposite trends in the dependence of the single charged pion ratio on the symmetry energy, pointing to the need for a better modeling of the reactions.  It is interesting that the sensitivity of SMASH to the symmetry energy is very weak. Since Coulomb force has not been implemented in SMASH, this suggests the possibility that pion production may reflect a subtle interplay of dynamics influenced by the Coulomb and symmetry energies.

\section{Summary and Outlook}

In summary, using the S$\pi$RIT Time Projection Chamber, we have measured charged pion multiplicities and their ratios for three Sn+Sn systems to $<4\%$ in multiplicity and $\approx2\%$ in ratio measurements. With radioactive beams, we are able to extend the $N/Z$ range between two Sn+Sn reactions by a factor of two compared to previous studies. By choosing systems with widely different $N/Z$ composition, but with similar Coulomb effects, we obtain the asymmetry dependence of single and double ratios. The precision reached in the data would allow a constraint on the symmetry energy if the factors contributing to the variation among the transport models were brought under control.  

Going beyond pion multiplicities and their ratios, calculations show that while low energy pions are affected by Coulomb and other dynamical effects, high energy pion spectra ratios show promise to allow constraining the density dependence of the symmetry energy as well as the momentum dependence of the isovector mean field potentials. We also have abundance of data for $Z=1$ and 2 particles. These light charged particles will be used to construct observables such as stopping, transverse and elliptical flows, to extract constraints of the isoscalar parameters in the transport models. In particular, $t/^3$He spectral ratios may provide some information on the symmetry potentials~\cite{Mor19}.

The results from the seven transport codes underscore the importance of understanding the current uncertainties in the prediction of pion observables related to the symmetry energy. The TMEP has developed benchmark calculations in a box to verify the technical implementation of pion production mechanisms and other aspects of a transport code. Future transport calculations should follow similar rigorous evaluations before comparison to data. Most codes adjusted their inputs to optimally reproduce the Au+Au data \cite{Rei07} which suffers from high experimental pion momentum thresholds. Availability of the accurately measured Sn+Sn momentum spectra should allow better adjustment of some of the input parameters to the transport codes. 

As long as the disagreement among the predictions of transport models for pion multiplicities and ratios is large, conclusions about the symmetry energy from the present comparison between calculations and data is precluded.  This likely explains the contradictory conclusions reached in previous pion production studies \cite{Xia09, Fen10, Xie13}. While it is beyond the scope of this letter to elucidate the origin of these discrepancies, we can comment on their possible causes. Based on the box comparison studies, for the seven codes that participate in the present work, the treatment of the collision term, the Pauli principle, the pion production mechanism, and the choice of the nucleonic isoscalar potentials appear to be under control. However, the mean-fields for pion and $\Delta$ are inadequately constrained and can strongly influence the energy available for subthreshold pion production. They are also relevant to the EoS and thermal transport properties of neutron stars at $\rho\sim2\rho_0$ \cite{Dra14,Jli19,For20}, beyond their relevance to the interpretation of pion production in the present experiment. 

Previous experience of constraining the symmetric matter EoS also illustrates how such properties of dense matter can be probed with an appropriately chosen set of experimental observables \cite{Dan02}. Calculations predict how the momentum dependencies of the nucleonic mean fields can be constrained by comparing neutron and proton energy spectra \cite{Cou16,Mor19,Zha14}.  We will need additional data that can be combined with the present data to obtain independent constraints on the local and non-local isovector mean field potentials at supra-saturation densities and on the production of pions and deltas in matter in the vicinity of twice saturation density. 

Measurements of pion spectra and pion production at different incident energies can test assumptions regarding the $\pi$ and $\Delta$ mean field potentials and study the $\Delta$ production threshold effects, and their impact on the pion production mechanism \cite{Tsa17, Hon14}. Sensitivity studies following \cite{Coz16,Coz17, Son15, Zha17, Fer05} may identify other areas where further improvements in the description of pion production in transport models are required.

\section{Acknowledgements}

The authors would like to thank the operation staff of the RIBF for producing high intensity $^{238}$U and $^{124}$Xe primary beams in order to produce the $^{132}$Sn, $^{124}$Sn, $^{108}$Sn and $^{112}$Sn secondary beams necessary for this study. This work was supported by the U.S. Department of Energy, USA under Grant Nos. DE- SC0014530, DE-NA0002923, DE-FG02-93ER40773, DOE DE-FG02-93ER40773, DE-SC0019209, DE-SC0015266, DE-AC02-05CH11231, US National Science Foundation, United States Grant No. PHY-1565546, the Robert A. Welch Foundation, United States (A-1266 and A-1358), the Japan Society for the Promotion of Science (JSPS) Grant-in-Aid for Scientific Research (KAKENHI) grant Nos. JP24105004, JP18H05404, JP17K05432 and JP19K14709, the National Research Foundation of Korea under grant Nos. 2016K1A3A7A09005578, 2018R1A5A1025563, 2013M7A1A1075764 the Polish National Science Center (NCN), Poland, under contract Nos. UMO-2013/09/B/ST2/04064, UMO-2013/-10/M/ST2/00624, the Deutsche Forschungsgemeinschaft (DFG, German Research Foundation) under Germany's Excellence Strategy – EXC-2094 – 390783311 (ORIGINS), SFB1245, the BMBF via Project No. 05P15RDFN1, EXC-2094-390783311, National Natural Science Foundation of China under Grants Nos. 11375094, 11079025, 11847315, 11875125, 11505, 11947410, 11922514, Croatian Science Foundation under projects Nos. 1257 and 7194. The computing resources for data analysis are provided by the HOKUSAI-GreatWave system at RIKEN, the Institute for Cyber-Enabled Research (ICER) at Michigan State University, and the EMBER cluster at the NSCL.

\end{document}